\documentclass{aastex6}

\fullcollaborationName{The Friends of AASTeX Collaboration}

\begin{document}


\title{Imaging and analysis of neon nova V382 Vel shell \footnote{Based on observations obtained at the Southern Astrophysical Research (SOAR) telescope, which is a joint project of the Minist\'erio da Ci\^encia, Tecnologia, Inova\c{c}\~ao e Comunica\c{c}\~oes (MCTIC) da Rep\'ublica Federativa do Brasil, the U.S. National Optical Astronomy Observatory (NOAO), the University of North Carolina at Chapel Hill (UNC), and Michigan State University (MSU).}}


\author{Larissa Takeda\altaffilmark{1} and Marcos Diaz\altaffilmark{2}}
\affil{IAG, Universidade de S\~ao Paulo \\
Rua do Mat\~ao, 1226 \\
Sao Paulo, SP 5508-900, Brazil}







\altaffiltext{1}{larissa.takeda@usp.br}
\altaffiltext{2}{marcos.diaz@iag.usp.br}

\begin{abstract}
We present the detection and imaging of the spatially resolved shell of nova V382 Vel with SOAR adaptive optics module (SAM). The shell was observed in narrow band filters H$\alpha$ and [O III] 5007\AA\, revealing different structures in each filter. The shell's angular diameter obtained was $9.9$ arcsec, equivalent to $2.8 \times 10^{17}$ cm, using the distance of $1.79$ kpc obtained by Gaia mission. The upper limit for total shell mass derived from recombination lines is $M_{s} = 1.4 \times 10^{-4}$ M$_{\odot}$. Our photoionization models indicate an accretion disk with $T_d=60,000$ K and L=$10^{36}$ erg/s as main ionizing source. 
\end{abstract}




\section{Introduction} \label{sec:intro}
Novae are a sub-type of cataclysmic variables that present at least one recorded high-amplitude eruption. Nova systems are composed of a white dwarf (primary star) and a less evolved secondary star (usually main sequence or red giant). The secondary transfers mass to the primary through an accretion disk or column. When the accreted surface layer achieves the necessary pressure to start hydrogen burning, the gas is ignited to a non-stable CNO cycle runaway \citep{Warner}. When in quiescence, the nova spectra are dominated by the H and He recombination lines of the disk, which contribute to novae unresolved spectra as soon as the accretion is reestablished. In order to obtain information about the host white dwarf and the physical properties of the eruption and of the hot remnant, we need to observe the emission and kinematics of the ejected gas itself. 

Among the few hundreds of detected novae, there is a select group of $\sim 20$ objects classified as neon novae. These are believed to be composed by O-Ne-Mg white dwarfs, which would explain the overabundance of neon, since this element is not produced in the eruption nucleosynthesis \citep{Williams1991}. From this group, only 5 objects have had their shells resolved, namely: V1974 Cyg, V351 Pup and QU Vul with HST observations \citep{Paresce,Downes,Krautter}, V1494 Aql with Russian BTA Telescope \citep{Barsukova} and V382 Vel with SOAR-SAMI (this paper) and SALT \citep{Tomov}, independently. In contrast, there are $26$ resolved nova shells, in total, available in the literature \citep{Camargo}. 

Neon novae occur under high-gravity regimes, with massive O-Ne-Mg white dwarfs that may be close to or approaching the Chandrasekhar's limit. O-Ne-Mg cores may be formed by different processes: single stars of initial mass between 8 and 12 M$_{\odot}$ \citep{Nomoto1984}, in close binary systems with mass transfer \citep{Isern}, and by the merging of two CO cores \citep{Nomoto1985}. The study of O-Ne-Mg white dwarfs in binary systems is important to constrain the physical conditions in which they occur, and the possibilities of those systems  becoming supernovae \citep{Nomoto1987}. One key information needed to understand neon novae outbursts and the evolution of O-Ne-Mg white dwarfs is the accurate neon and oxygen abundances of each object. Nevertheless, these abundances have been often derived from simplistic models, usually considering uniform or symmetric mass distribution in the shell. For reliable abundances estimates, high spatial resolution 2-D spectroscopy is essential to build complete 3-D photoionization models \citep{Moraes}.

It is quite difficult to obtain spatially resolved spectra of nova shells due to small angular diameter and surface brightness constraints, combined with the presence of a bright central source. It may take some years after the eruption so that the gas will expand far enough to be properly observed. On the other hand, the central source becomes fainter with time, and its ionization decreases with time and distance within the shell, making it more difficult to observe the shell line emission. In this scenario, one technique that facilitates the resolution of structures in nova shells is narrow band imaging aided by adaptive optics (AO).

In this paper we use the SOAR adaptive optics module imager (SAMI) to map the shell of nova V382 Vel. This nova was discovered on May 22nd of 1999 at $V = 3.1$ \citep{V382VelDisc}, making it one of the brightest novae ever detected. V382 Vel is a fast nova, with $t_{3} = 17.5$ days \citep{Liller}. \cite{Woodward} classified V382 Vel as a neon nova after the detection of [Ne II] $12.81$ $\mu$m in the IR spectra taken on July, 1999. X-rays data taken at the early eruption phases suggest an internal shock, caused by two consecutive ejections or by the interaction between the accelerated gas and the material from the expanded pseudo-photosphere. After 6 months of eruption, the X-rays observations presented a supersoft spectrum \citep{Mukai, Orio}.

\section{SOAR-SAMI observations} \label{sec:sami}
Continuum, $H\alpha$ and [O III] $\lambda 5007$ narrow band images were made of V382 Vel using SOAR imager with adaptive optics module (SAM) at $\Delta t = 5781$ days after eruption. The SAM delivered an image quality (FWHM) of $\sim 0.54"$ for H$\alpha$ and continuum filters and $\sim 0.60"$ for the [O III] one, making it possible to acquire a detailed emission map of the shell. We obtained 2 exposures of $1200$ s for H$\alpha$ and [O III] $5007$\AA\ filters and 2 exposures of $600$ s for y-Str\"{o}mgren (continuum) filter. Photometric standards LTT 1788 and LTT 3864 were used for flux calibration. 

\begin{figure}
\begin{center}
\includegraphics[scale=1.0]{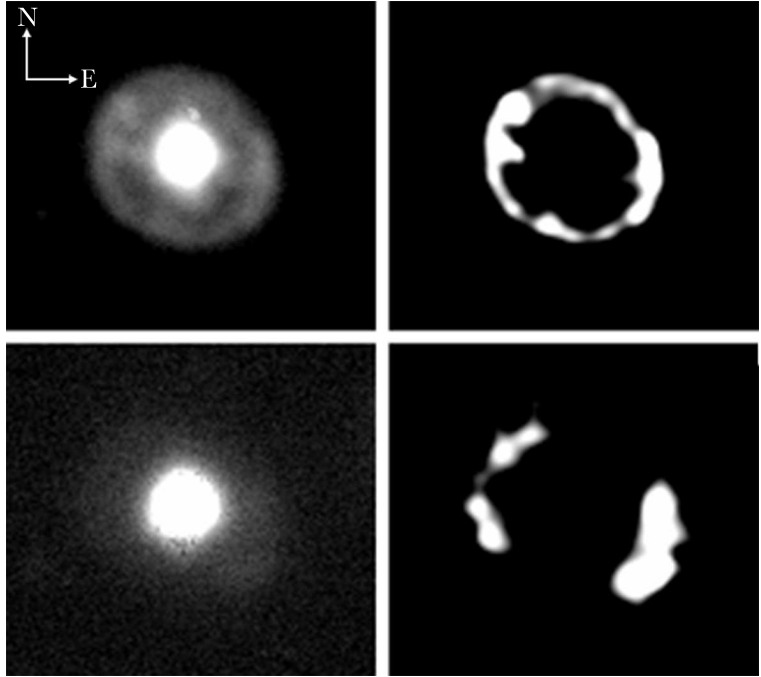}
\caption{Left: H$\alpha$ (top) and [O III] (bottom) calibrated images obtained by SOAR-SAMI of nova V382 Vel. Right: Reduced, deconvolved H$\alpha$ (top) and [O III] (bottom) shell images of V382 Vel with both continuum contribution and central source emission subtracted.}
\label{sami}
\end{center}
\end{figure}

Basic reduction was performed, applying bias and sky-flat corrections. LTT 3864 images were used for the flux calibration, while LTT 1788 images were only used to crosscheck the sensitivity ratios between filters, since they were taken through thin cirrus clouds. We combined the images for each filter and performed aperture photometry and Richardson-Lucy deconvolution with IRAF tools \citep{Iraf}. As the PSF in H$\alpha$ and y-Str\"{o}mgren images were narrower than in [O III] image, we degraded them in order to match the [O III] image PSF. The emission of the central source in the [O III] $5007$\AA\ band corresponds only to the continuum emission, as neither the disk nor the nova pseudo-photosphere are expected to produce forbidden transition's lines. Therefore, we fitted a Gaussian function on the PSF of the central source and subtracted it from the image. The result is the [O III] $5007$\AA\ isolated contribution from the shell. The disentanglement of the central source continuum and shell contributions for H$\alpha$ is more complex, because all structures emit strong hydrogen lines. Using the flux of y-Str\"{o}mgren image, we isolated the H$\alpha$ line emission in our image. Then, we fitted a Gaussian function to the central source PSF and subtracted it from the image, leaving only the H$\alpha$ contribution from the shell. The resulting images for H$\alpha$ and [O III] $5007$\AA\ emission from the shell are displayed in figure \ref{sami}.

\section{V382 Vel shell}
\subsection{Angular diameter}
\begin{figure}
\begin{center}
\includegraphics[scale=0.7]{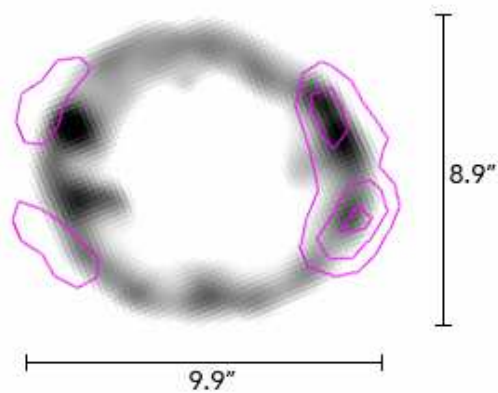}
\caption{Deconvolved shell image in H$\alpha$ filter with overlapping [O III] contours. The shell major and minor axes are displayed.}
\label{overplot}
\end{center}
\end{figure}

H$\alpha$ image reveals a thin and roughly spherical shell, with a few resolved clumps. The major and minor axes are $9.9$ arcsec and $8.9$ arcsec, respectively, with a mean value of $9.4$ arcsec (see figure \ref{overplot}). \citet{Tomov} measured an angular diameter of $12$ arcsec for the shell from their H$\alpha+$[N II] narrow band imaging with Fabry-Perot Robert Stobie Spectrograph at SALT.

[O III] $5007$\AA\ image shows a bipolar structure instead of a complete shell. This structure has a slightly larger angular diameter, of $10.9$ arcsec, and the clumps are not completely consistent with the H$\alpha$ ones. The difference in size, and perhaps the clumps mismatch, may be due to the density distribution associated to expansion velocity gradients. The H$\alpha$ emission occur in denser regions than the [O III], and the density is lower at the outer part of the shell. Figure \ref{overplot} also shows the H$\alpha$ emission map with the [O III] emission as a contour plot, stressing the difference in shell size. 

Most novae with high expansion velocities present spherically symmetric shells, and the ones with low expansion velocity may present bipolar structures \citep{Camargo}. The shell morphology of V382 Vel may be also related to the occurrence of two successive eruption peaks, indicated by the x-ray detection of a internal shock soon after outburst maximum \citep{Mukai, Orio}. If there was a low velocity eruption followed by a high velocity one, the first would originate the bipolar structure seen in [O III] profile and the second may have originated the almost spherical shell seen in H$\alpha$ band. 

We do not have information about the inclination of the system, and we do not know if the bipolar structure is in the orbital plane or perpendicular to the binary orbit. If the [O III] structure is perpendicular to the orbital plane, it may be an indication of the presence of the disk, generating an anisotropic ionizing field, as for the case of HR Del \citep{MoraesHRDel} and V723 Cas \citep{TakedaV723Cas}. 

\subsection{Expansion velocity and parallax distance}
Our analysis and models consider a spherical shell with radius of $4.95$ arcsec, assuming the [O III] emission as bounder. Different values of expansion parallax distance were derived from several author's expansion velocities. The expansion velocities for V382 Vel were measured as the HWZI of the emission lines, thus the values found in the literature broadly vary, possibly because they are highly dependent on the continuum S/N. In some cases, other broadening effects unrelated to the gas group velocity may be present. The velocity of $\sim 1200$ km/s found by\citet{Augusto} during the first 800 days after eruption, yields a distance of $d = 810$ pc. \citet{Tomov} obtained a distance of $800$ pc, using their velocity measurement of $1800$ km/s and their angular diameter of $12.0$ arcsec. Using the same velocity of $1800$ km/s and the shell angular diameter value obtained by SOAR-SAMI, we obtain a distance of $1214$ pc.

Gaia DR2 \citep{Gaia2016,Gaia2018} contain a parallax value of $0.56$ mas with error of $\sim 10\%$ for V382 Vel, leading to a distance of $1.79$ kpc. In order to match such a distance with our angular radius value, the average expansion velocity should be around $2600$ km/s. Although higher than the velocities found by \citet{Augusto} and \citet{Tomov}, this value is still compatible with the speed class of V382 Vel (fast nova). The distance of $1.79$ kpc is also consistent with the values derived by \citet{DellaValle} and \citet{HachisuNeon}, of $1.7$ kpc (from MMRD) and $1.6$ kpc (from distance-reddening relation), respectively. \citet{ShoreV382Vel} gave a higher estimate of $2.5$ kpc using observational constraints as the presence of interstellar lines in the spectra and maximum luminosity. In the following simulations, a distance $d=1.79$ kpc and an outer shell radius $r = 1.4 \times 10^{17}$ cm were assumed.

\subsection{Line fluxes and dereddening}
As commented before, we calibrated V382 Vel data using the spectrophotometric standard LTT 3864. We also corrected the fluxes from interstellar extinction using $E(B-V) = 0.38$, taken from the IRSA Galactic Dust Reddening and Extinction calculator, which uses dust reddening from \citet{SandF}. \citet{DellaValle} derived $E(B-V)=0.1$ using the equivalent width from interstellar Na lines in their optical spectra. As these line EWs may be close to saturation, we decided to use the value from the galactic dust map. The final fluxes obtained for the shell emission lines are $f_{H\alpha}=9.9\times 10^{-16}$ erg/s/cm$^2$ and $f_{[O III]}=7.7\times 10^{-16}$ erg/s/cm$^2$. The flux ratio between [O III] and H$\alpha$ of $0.78$ is relatively high among evolved neon nova shells. Analyzing the optical spectrum of \citet{Tomov}, the ratio between [O III] and H$\alpha$ fluxes is roughly $0.1$, indicating that their spectrum is already dominated by the bright hydrogen lines from the accretion disk.

We have estimated that the flux loss due to the broad line wings and the limited width of the narrow band filters is less than $10\%$ for both bands, using previous spectroscopic observations \citep{Tomov}. Such loss is not relevant when compared to other uncertainties in our models.

\subsection{Deprojection of the shell H density}
\begin{figure}
\begin{center}
\includegraphics[scale=0.8]{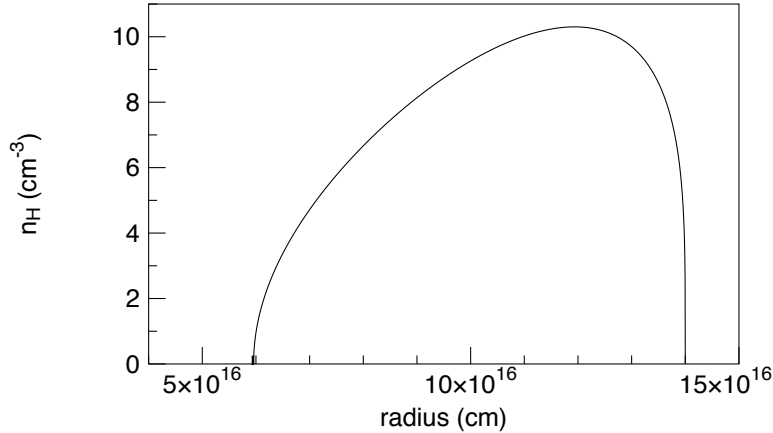}
\caption{Diffuse hydrogen density radial profile derived from Abel inverse transform.}
\label{abel}
\end{center}
\end{figure}

Even though our images suggest that the shell of V382 Vel is not completely spherical, we do not have information about its eccentricity, due to the projection effects. Thus, in the following analysis we assume a clumpy structure added to a diffuse spherical component. Under the spherical simplifying assumption it was possible to use the projected recombination line emissivity to estimate a 3D mass distribution. First, we isolated the clumps from the diffuse gas, by measuring the median flux value as a function of the radius. In order to obtain a volume emissivity from the surface brightness, we used an Abel inverse transformation \citep{Park}, with a polynomial fit. The volume emissivity distribution was then rescaled into a density distribution, assuming a a fully ionized gas at a trial temperature of $10,000$ K \citep{Osterbrock}. The density gradient (figure \ref{abel}) was smoothed at the borders to guarantee the convergence of photoionization calculations and then used to produce a 3D density map. 

The features on the residual image (obtained by the subtraction of the radial median flux from the H$\alpha$ image) were treated as clumps. We submitted the residual image to a 2D Fourier Transform in order to obtain a characteristic scale of the structures. The power spectrum indicated a predominance of components with diameter $d=9\times 10^{15}$ $cm$, a value chosen to define the clumps' size. The 3 brightest peaks coordinates in the residual image were used as the projected clumps centers while the positions along the line of sight were randomly chosen within the shell. For each clump, the integrated flux was scaled to the observed value, with a Gaussian radial gas density distribution. These clumps were added to the 3D hydrogen density map.

\subsection{Hydrogen mass}
Using the 3D shell H density distribution, combined with the estimated distance and shell radius, we derived a total hydrogen mass of $M_{shell}=6.2\times 10^{-5}$ M$_{\odot}$. This value leads to an estimate of the mean hydrogen density of $n_{H} \sim 15$ cm$^{-3}$. For a hydrogen mass fraction $X=0.46$, calculated from the abundances of table \ref{tab:inputpar}, the total ejected mass is $M_{shell}=1.4\times 10^{-4}$ M$_{\odot}$. This value should be considered an upper limit for the ejected mass, since the observed H$\alpha$ flux may have a contribution of [N II] lines due to the width of the filter. Our estimate is compatible with the hydrogen mass derived by \citet{ShoreV382Vel} of $M_{shell} = 4\times 10^{-4}$ M$_{\odot}$, using UV He II observed fluxes compared to photoionization predictions. These models assumed $d=2.5$ kpc, $E(B-V)=0.2$, two symmetric shell components with different densities and a covering factor of $0.8$. \citet{DellaValle} found a lower value of $M_{shell}=6.5\times 10^{-6}$ M$_{\odot}$, using the H$\alpha$ emission from year 2000 observations ($\Delta t \sim 17$ months), although it is not clear how the authors separated the shell emission from the central source. \citet{HachisuNeon} derived a $M_{shell}=4.8\times 10^{-6}$ M$_{\odot}$ at optical maximum in their light-curve models.

\section{Photoionization models}

We used RAINY3D photoionization models \citep{Moraes}, that are based on Cloudy \citep{Ferland}, to explore the properties of V382 Vel nova system. The input parameters are listed in table \ref{tab:inputpar} used with the geometry constraint from the 3D hydrogen density map previously described.

\begin{deluxetable}{cc}
\tablecaption{Input parameters \label{tab:inputpar}}
\tablehead{
\colhead{Parameter} & \colhead{Value}}
\startdata
dist & $d= 1.79$ $kpc^{[1]}$ \\
E(B-V) & $0.38^{[2]}$\\
$T_{CS}$ & $5.0-8.0$ $(\times10^4)$ $K^{[3]}$ \\
$log(L_{CS})$ & $35-36$ $erg/s^{[3]}$ \\
$r_{in}$ & $6.0\times10^{16}$ $cm^{[3]}$  \\
$r_{out}$ & $1.4\times 10^{17}$ $cm^{[3]}$ \\
$M_{H}$ & $1.4\times 10^{-4}$ $M_{\odot}^{[3]} $  \\
$log(n_{He}/n_{H})$ & $-0.6^{[4]}$  \\
$log(n_{C}/n_{H})$ & $-3.7^{[5]}$  \\
$log(n_{N}/n_{H})$ & $-2.8^{[5]}$ \\
$log(n_{O}/n_{H})$ & $-3.8^{[4]} - -2.6^{[5]}$  \\
$log(n_{Ne}/n_{H})$ & $-3.0^{[4]} - -2.7^{[5]}$  \\
$log(n_{Mg}/n_{H})$ & $-4.0^{[5]}$  \\
$log(n_{Al}/n_{H})$ & $-4.2^{[5]}$  \\
$log(n_{Si}/n_{H})$ & $-4.8^{[5]}$  \\
\enddata
\tablenotetext{1}{\citet{Gaia2016,Gaia2018}}
\tablenotetext{2}{\citet{SandF}}
\tablenotetext{3}{This paper.}
\tablenotetext{4}{\citet{Augusto}}
\tablenotetext{5}{\citet{ShoreV382Vel}}
\end{deluxetable}

The central source luminosity and temperature inputs were based on average values for novae $\sim 10$ years after eruption, since there were no X-rays data taken at the same epoch of our observations. In 2000, \citet{Ness2005} found an integrated blackbody luminosity of $2\times 10^{36}$ erg/s by the time the hydrogen burning had already turned off - a value used as an upper limit in our models. These authors also found a central source temperature of $3\times10^{5}$ K, but it is expected to have greatly declined by the epoch of SOAR-SAMI observations. 

\begin{figure}
\begin{center}
\includegraphics[scale=1.0]{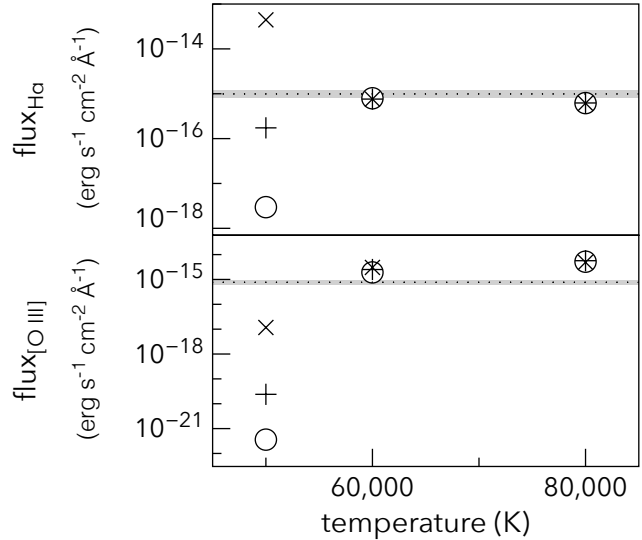}
\caption{Rainy3D model integrated line fluxes as a function of the ionizing source temperature. The circles correspond to an ionizing source luminosity of $10^{35}$ $erg/s$, the plus signs to $10^{35.5}$ $erg/s$ and the `x' marks to $10^{36}$ $erg/s$, respectively. The dotted lines indicate the observed line fluxes.}
\label{rainy_models}
\end{center}
\end{figure}

The first RAINY3D models were made without the condensations in order to limit the input parameters range, because the clumpy models are much more time-consuming. The results are displayed in figure \ref{rainy_models} and suggest an ionizing source of $T=60,000$ K, but the H$\alpha$ fluxes are lower than expected while the [O III] fluxes are higher. As in the case of nova V723 Cas \citep{TakedaV723Cas}, one possible solution that could better model V382 Vel is the anisotropy of the ionizing radiation field caused by an accretion disk. It is reasonable to assume that the ionizing source is dominated by the accretion disk after more than $15$ years after V382 Vel eruption. The luminosities and temperatures used in our models are in fact compatible with accretion disks of $ 5\times 10^{-8} \leq \dot{M} \leq 1\times 10^{-7}$ M$_{\odot}/year$. 

Therefore, we ran the clumpy RAINY3D photoionization models including a standard model accretion disk with $T_d=60,000$ K as ionizing source. We assumed a typical value of M$_{WD}=1.1$ M$_{\odot}$ for an O-Ne-Mg white dwarf and a disk radius of $80\%$ of the primary Roche-Lobe radius. The total luminosity of the disk was fixed in $10^{36}$ erg/s, because the anisotropic models usually require higher luminosities than isotropic ones, in order to to reproduce the total fluxes. The inclination of the disk is unknown. However, bipolar structures as the one presented by [O III] emission tend to be perpendicular to the disk \citep{TakedaV723Cas,HarveyGKPer}. Accretion disks parallel and perpendicular to the bipolar structure were modeled and compared. In the first case, the ionizing flux towards the bipolar structure is the lowest and in the second case, the highest.

\begin{figure}
\begin{center}
\includegraphics[scale=0.60]{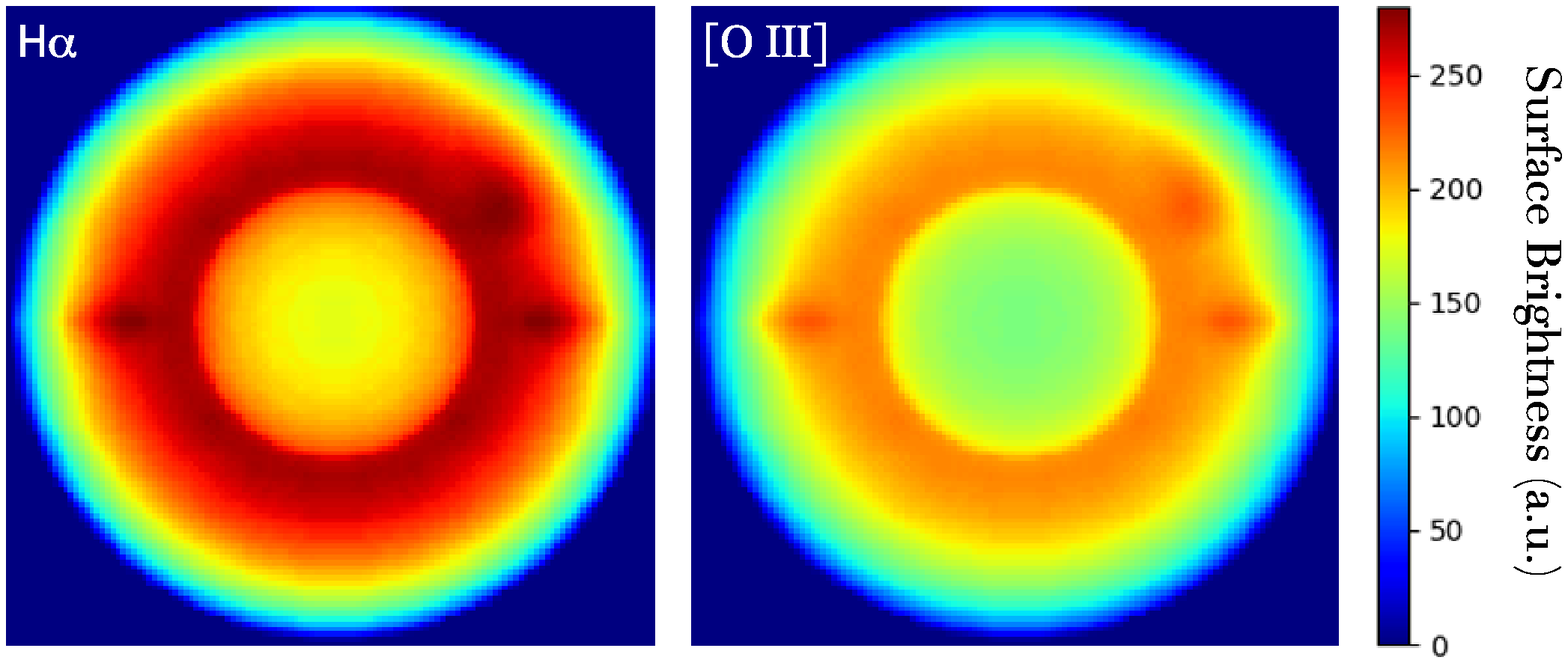}
\includegraphics[scale=0.60]{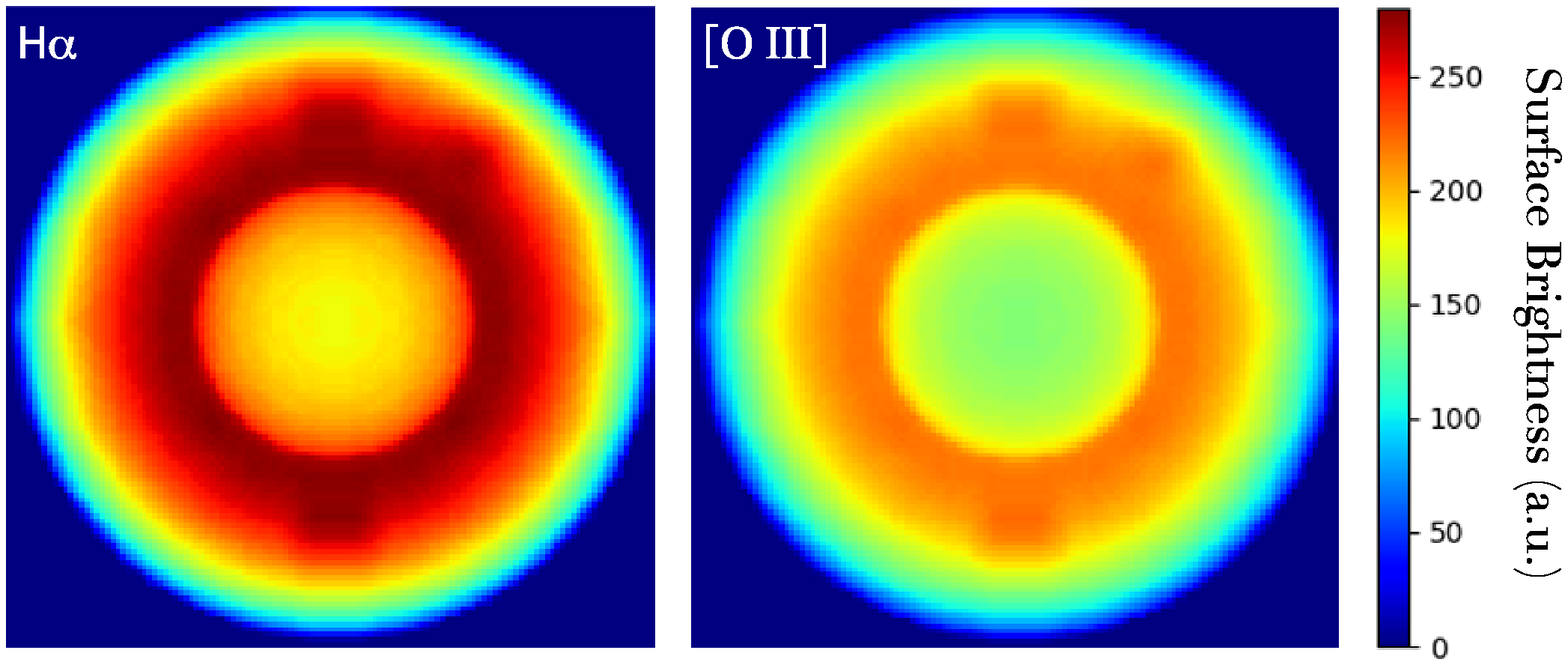}
\caption{RAINY3D projected model emissivity maps for H$\alpha$ and [O III], for bipolar structure parallel (top) and perpendicular (bottom) to accretion disk, which is aligned horizontally. The surface brightness values are displayed in arbitrary units.}
\label{emmap}
\end{center}
\end{figure}

Our best fit model for a parallel disk is displayed in figure \ref{emmap}, as a projection of the resulting 3D emissivity maps for H$\alpha$ and [O III]. A logarithmic arbitrary intensity scale is used to emphasize faint structures. The total modeled fluxes of H$\alpha$ and [O III] are f$_{H\alpha}$=$4.6\times 10^{-15}$ erg/s/cm$^2$ and $1.3\times 10^{-15}$ erg/s/cm$^2$, which are 4.6 and 6.0 times higher than observed values, respectively. The best fit model for a perpendicular disk is also displayed in figure \ref{emmap}, with total modeled fluxes of f$_{H\alpha}$=$2.3\times 10^{-15}$ erg/s/cm$^2$ and $6.5\times 10^{-16}$ erg/s/cm$^2$, with factors of 2.3 and 0.7 higher than the observations. The clumps positions, size and shape do not match exactly, but the model emissivities reproduce the observations in the sense that [O III] clumps end further from the central source than H$\alpha$ ones. Spatially resolved spectroscopy including transitions in a wide ionization range is needed to improve the models. For these best fit models, with the disk being parallel or perpendicular to the bipolar structure, the oxygen abundance is revised as $log(n_{O}/n_{H}) = -3.8$, a value which is consistent with that obtained by \citet{Augusto}. \citet{ShoreV382Vel} found a higher value for O abundance, using photoionization models based on UV spectra, with a higher distance value of $2.5$ kpc and a two component spherical shell with different filling factors. A precise derivation of the O and Ne abundances can elucidate the stellar evolution in the binary system and the dragging and mixing processes at the surface of the white dwarf during eruption.

Given that we only have data for 2 transitions, our models are poorly constrained to provide an unique solution for the shell emission. The differences between [O III] and Balmer lines are likely due to a combination of local conditions that produce different collisional excitation and ionization of O$^{+}$, including the proposed anisotropic radiation. We are able to rule out any collisional de-excitation effects to [O III] given the low densities involved.

\section{Discussion} \label{sec:discussion}

The use of adaptive optics made the detection of clumps in V382 Vel shell possible, and the 2D Fourier Transform gave a rough estimate of their scale. However, it is plausible that the structures we detect as clumps are actually large groups of smaller clumps, smaller than we can detect with our AO resulting image quality. ALMA observations of nova V5668 Sgr presented dispersed structures of $\sim 10^{15}$ cm that were interpreted as large unique structures when observed with lower resolution instruments \citep{ALMA}. The misinterpretation of the clumps size and filling factor can affect the H densities, since we considered the radial average emission as diffuse component.

The asymmetries in nova shells have been extensively detected in the past decades, but they are usually attributed to anisotropic mass distributions, driven by ejection processes, post-maximum winds and interaction of the expanding gas with the companion star and pre-existing circumstellar material \citep{Walder}. In our models, we inserted symmetric clumps, but due to the anisotropic ionizing field they seem asymmetric in the emissivity maps. This may be seen as an indication that clumps and mass distributions are not necessarily asymmetric, even if the projected observed fluxes are. The actual contribution of each factor for the observed asymmetries could eventually be studied by joining spatially resolved spectroscopy, 3D hydrodynamic simulations of clump evolution and 3D photoionization models.

In our analysis, we derived a total shell mass of $\sim 1.4 \times 10^{-4}$ M$_{\odot}$, that lies on the upper limit of nova shell masses, especially for fast and/or neon novae. Considering that the X-rays detection of a shock is due to a second eruption with larger expansion velocities, the ejected mass could be higher than expected for a single eruption. The high ejected mass found favors the scenario where O-Ne-Mg novae are not progenitors of neutron stars via accretion-induced collapse \citep{Nomoto}, as their mass loss overcomes the accretion.

V382 Vel belongs to the small group of confirmed neon novae. Among the objects of this group, only other 4 novae have had their shells resolved: V1494 Aql, V1974 Cyg, V351 Pup and QU Vul. As for V382 Vel, V1974 Cyg imaging also presented a inhomogeneous ring, or thin shell, and a bipolar structure \citep{Paresce,Paresce1995}. These structures appear differently for distinct ionization lines, a possible consequence of the presence of the accretion disk as ionizing source. \citet{Krautter} derived a high ejected mass for V1974 Cyg, of $2.0\times 10^{-4}$ M$_{\odot}$, based on data from \citet{Woodward1995}, while \citet{HachisuNeon} found the value of $1.3\times 10^{-5}$ M$_{\odot}$ through light-curve models. Images from \textit{HST} of QU Vul and V351 Pup also show clumpy rings \citep{HST,Krautter}. \citet{SaizarQUVul} restricted the total ejected mass of QU Vul to the interval $1.0\times 10^{-4}$ M$_{\odot}$ to $3.5\times 10^{-3}$ M$_{\odot}$. \citet{HachisuNeon} derived an ejected mass of $2.5\times 10^{-5}$ M$_{\odot}$ by modeling QU Vul light-curve. For V351 Pup, \citet{Wendeln} found a total ejected mass of $6.3\times 10^{-6}$ M$_{\odot}$, \citet{SaizarV351Pup} found $2.0\times 10^{-7}$ M$_{\odot}$ and \citet{HachisuNeon} found $2.0\times 10^{-5}$ M$_{\odot}$. \citet{Barsukova} describe V1494 Aql shell as a thin spherical shell or a ring.

The observed shells of neon novae seem to present common morphological features. However, the ejected masses vary over a wide range and in some cases the masses derived by different authors broadly diverge. The major problems in estimating the shell masses are the actual gas distributions or proper filling factors, and the errors in distances. Gaia mission may solve the last one, but for the first spatially resolved spectroscopy or multi-wavelength high-resolution imaging is required. Only then we will be able to characterize neon novae as a group and connect the observations to the theoretical predictions.

\section{Conclusions}

The SOAR-SAMI observations presented a detailed view of a thin, roughly spherical shell in H$\alpha$ filter and a bipolar structure in [O III] filter. Both images present clumps, although they are not aligned in both filters. We derived a shell radius of $1.4\times 10^{17}$ cm, directly from our images and Gaia parallax distance, and a upper limit for total shell mass of $\sim 1.4 \times 10^{-4}$ M$_{\odot}$. The H density distribution, the ionizing source temperature of $T_d=60,000$ K, a total central source luminosity of L=$10^{36}$ erg/s and the oxygen abundance in the ejecta were constrained from our photoionization models. These models also suggest that the gas is ionized by the re-established accretion disk.

\acknowledgments

We thank FAPESP for the support under grant 2014/10326-3 and CNPq funding under grant \#305657. This research has made use of the NASA/ IPAC Infrared Science Archive, which is operated by the Jet Propulsion Laboratory, California Institute of Technology, under contract with the National Aeronautics and Space Administration. This work has made use of data from the European Space Agency (ESA) mission {\it Gaia} (\url{https://www.cosmos.esa.int/gaia}), processed by the {\it Gaia} Data Processing and Analysis Consortium (DPAC, \url{https://www.cosmos.esa.int/web/gaia/dpac/consortium}). Funding for the DPAC has been provided by national institutions, in particular the institutions participating in the {\it Gaia} Multilateral Agreement.



\vspace{5mm}
\facilities{SOAR(SAMI)}

\software{IRAF, Cloudy, RAINY3D}

\bibliographystyle{aasjournal}
\bibliography{bibliografia}

\allauthors

\listofchanges

\end{document}